\documentclass[sigconf, nonacm]{acmart}

\usepackage{ragged2e} 
\usepackage{svg}
\usepackage{tabularx}
\usepackage{enumitem}
\usepackage{multirow} 

\begin{document}
\title{VectraFlow: Long-Horizon Semantic Processing \texorpdfstring{\\}{ }over Data and Event Streams with LLMs}

\author{Shu Chen}
\authornote{Both authors contributed equally to this research.}
\affiliation{%
  \institution{Brown University}
}
\email{shu_chen@brown.edu}

\author{Junhan Liu}
\authornotemark[1]
\affiliation{%
  \institution{Brown University}
}
\email{junhan_liu@brown.edu}

\author{Deepti Raghavan}
\affiliation{%
  \institution{Brown University}
}
\email{deepti_raghavan@brown.edu}

\author{Ugur Cetintemel}
\affiliation{%
  \institution{Brown University}
}
\email{ugur_cetintemel@brown.edu}

\begin{abstract}
Monitoring continuous data for meaningful signals increasingly demands \emph{long-horizon, stateful} reasoning over unstructured streams. However, today's LLM frameworks remain stateless and one-shot, and traditional Complex Event Processing (CEP) systems, while capable of temporal pattern detection, assume structured, typed event streams that leave unstructured text out of reach. We demonstrate \textit{VectraFlow}, a semantic streaming dataflow engine, to address both gaps. VectraFlow extends traditional relational operators with LLM-powered execution over free-text streams, offering a suite of \textit{continuous semantic operators}---filter, map, aggregate, join, group-by, and window---each with configurable throughput--accuracy tradeoffs across LLM-based, embedding-based, and hybrid implementations. 

Building on this, a \textit{semantic event pattern operator} lifts complex event processing to unstructured document streams, combining LLM-based event extraction with NFA-based temporal rule matching for stateful reasoning over sequences of semantic events. 

In this demonstration, users will interact with VectraFlow's live query interface to compose semantic pipelines over clinical document streams. Attendees will compile natural language intents into executable operator graphs, inspect intermediate stateful outputs, and observe end-to-end temporal pattern detection, from raw text to matched event cohorts.
\end{abstract}
 
\maketitle

\pagestyle{plain} 

\setlength{\textfloatsep}{4pt}

\section{Introduction}

Large language models (LLMs) have become the de facto tool for interpreting unstructured data, yet most LLM pipelines remain stateless and episodic, evaluating isolated prompts over static corpora without support for persistence or reasoning that unfolds over time. Many applications, however, require signals that emerge across sequences of events rather than single documents. Detecting a deteriorating patient, escalating compliance violations, or multi-step fraud patterns demands long-horizon, stateful reasoning over unstructured streams---a capability that remains beyond both one-shot LLM prompting and even today’s agentic models.

Existing systems address only part of this challenge. LLM-based frameworks such as Palimpzest~\cite{liu2025palimpzest}, LOTUS~\cite{patel2025semanticoperators}, and DocETL~\cite{shankar2025docetl} introduce semantic operators for unstructured data, but operate in batch or one-shot settings with no cross-record state. In contrast, complex event processing systems~\cite{wu2006sase, agrawal2008efficient, flinkcep, esper} enable long-horizon temporal reasoning but assume structured data. Neither approach suffices when events must first be \emph{inferred} from unstructured text before being \emph{reasoned over}.

In this demo, we will showcase \textit{VectraFlow}~\cite{vectraflow-cp,vectraflow-cidr}, a semantic streaming dataflow engine that bridges this gap by unifying continuous semantic processing with complex event pattern detection in a single framework. Analogous to how continuous queries extend relational operators to unbounded streams, VectraFlow lifts LLM-based semantics into continuous, pipeline operators. It extends both relational processing and event pattern detection with LLM-augmented execution over unstructured streams, enabling queries that span from record-level semantic interpretation to long-horizon temporal pattern matching. We will showcase these features via:
\begin{itemize}[leftmargin=*, itemsep=1pt, topsep=2pt]
  \item An \emph{interactive, end-to-end demonstration} enabling users to compose and observe semantic pipelines over unstructured streams. A natural language interface compiles queries into executable operator graphs with live visibility into intermediate states.
  
  \item A suite of \emph{continuous semantic operators} (e.g., filter, aggregate, window) supporting tunable throughput--accuracy tradeoffs via LLM-based, embedding-based, and hybrid execution strategies.
  
  \item A \emph{semantic pattern operator} that extends complex event processing to unstructured text, fusing LLM-based event extraction with NFA-based matching for long-horizon temporal reasoning.
\end{itemize}

\section{VectraFlow Overview}
VectraFlow is a semantic streaming dataflow engine that extends traditional relational operators and complex event processing with LLM-backed execution over unstructured streams. 

\begin{figure*}[htbp]
    \centering
    \includegraphics[width=\linewidth]{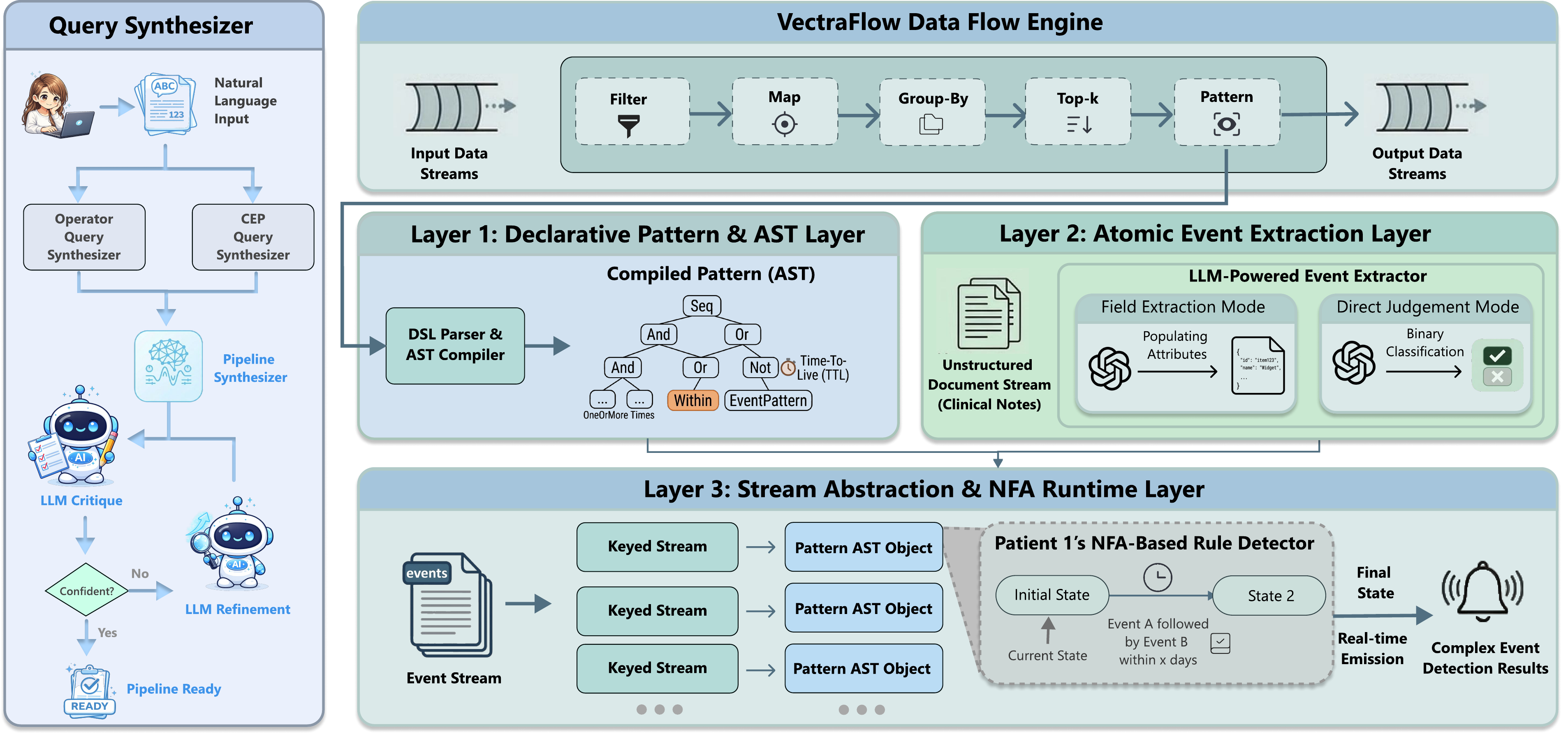}
    \caption{VectraFlow Data and Event Processing Architecture.}
    \label{fig:system_arch}
\end{figure*}


\noindent\textbf{Core Architecture.} Figure~\ref{fig:system_arch} illustrates the overall VectraFlow architecture, which is built on three layers.
At the foundation, the \emph{streaming dataflow engine} models computation as a directed acyclic graph (DAG) of operators through which records flow continuously from sources to sinks.
Above this layer, the \emph{LLM-based semantic operator layer} extends traditional relational operators with LLM-powered execution over unstructured text. This layer also incorporates \texttt{sem\_pattern}, which encapsulates event extraction and temporal rule matching behind a single composable operator. Finally, the \emph{natural language layer} bridges the gap between user intent and executable pipelines through an agentic loop of structured critique, automatic repair, and targeted user clarification.

\noindent\textbf{Continuous Semantic Operators.} VectraFlow provides a suite of semantic operators that extend their traditional relational counterparts with LLM-powered execution over unstructured text.
Operators such as Semantic Filter, Map, Aggregate, Join, and Group-By are direct analogs of their relational counterparts, preserving familiar query semantics over free-text streams.
Beyond these analogs, VectraFlow introduces operators designed for dynamic, open-ended streams: Semantic Window adjusts boundaries on topic or sentiment shifts, Semantic Group-By allows categories to emerge and dissolve over time, and Continuous RAG adapts retrieval context as query scope evolves. Table~\ref{tab:semantic-ops} provides details on some of VectraFlow's key operators (full set of operators described in~\cite{vectraflow-cp}).

\begin{table}[t]
\centering
\footnotesize
\setlength{\tabcolsep}{3pt}
\renewcommand{\arraystretch}{1.05}
\caption{Sample continuous semantic operators.}
\label{tab:semantic-ops}

\begin{tabularx}{\linewidth}{
@{}
p{1.4cm}
>{\RaggedRight\arraybackslash\hsize=0.8\hsize}X
>{\RaggedRight\arraybackslash\hsize=1.2\hsize}X
@{}}
\toprule
\textbf{Operator} & \textbf{Semantics} & \textbf{Implementation} \\
\midrule

\texttt{sem\_window} &
Adaptive segmentation based on topic or sentiment shifts. &
Pairwise similarity, rolling summaries, or embedding-based clustering. \\

\texttt{sem\_groupby} &
Online grouping by meaning with evolving categories. &
LLM assignment/creation with periodic merge--split refinement; embedding clustering + LLM labels. \\

\texttt{cont\_rag} &
Continuously updates retrieval context as query scope evolves. &
Adaptive prompting (unified or decomposed) with LLM or embedding retrieval. \\

\texttt{sem\_pattern} &
Temporal pattern detection over events extracted from text streams. &
LLM-based event extraction + NFA rule matching (single pass). \\

\bottomrule
\end{tabularx}
\end{table}

\begin{figure}[t]
    \centering
    \includegraphics[width=\linewidth]{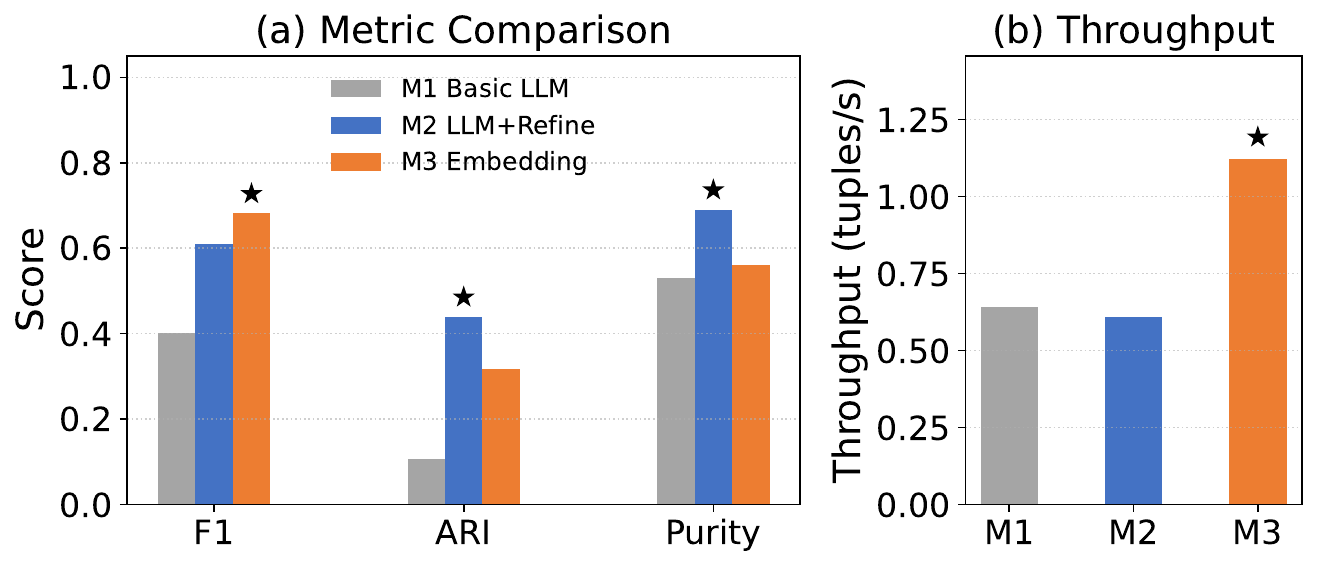}
    \caption{Semantic group-by implementations on the MiDe22 dataset. Left: comparison across F1, ARI, and Purity ($\star$ marks the best). Right: throughput in tuples/s.}
    \label{fig:sem-groupby}
\end{figure}

\noindent \textbf{Experimental Results.} VectraFlow operators typically support LLM-based, embedding-based, or hybrid implementations to trade off semantic fidelity against latency and token cost. Here we illustrate this tradeoff using \texttt{sem\_groupby} as a representative example~\cite{vectraflow-cp}, where we compare the alternative implementations on a subset of MiDe22~\cite{toraman2024mide22}, reporting F1, ARI, Purity, and throughput (tuples/s) in Figure~\ref{fig:sem-groupby}.
The \emph{LLM with Refinement (M2)} method periodically issues an additional refinement prompt every 10 tuples.
\emph{Embedding-based (M3)} grouping is fast and achieves high item-level F1, but its over-segmentation produces fragmented events.
\emph{Basic LLM (M1)} offers moderate coherence and competitive speed, whereas \emph{LLM with Refinement (M2)} improves cluster coherence metrics at the cost of lower throughput, making it the preferred choice when preserving event structure is the primary objective.

\section{Semantic Pattern Operator}

Real-world analytics often demand long-horizon temporal reasoning over sequences of events. Traditional CEP systems~\cite{wu2006sase, akdere2008plan, aurora, flinkcep, esper, snowflake-match-recognize} excel in this, but they rely on structured, explicitly typed event streams, which makes them essentially unable to detect events that are hidden within unstructured free text. Conversely, while semantic relational operators effectively interpret unstructured text, they process records in isolation and cannot express cross-record temporal constraints, such as sequential ordering or time-windowed conditions. To bridge this gap, VectraFlow introduces the \texttt{sem\_pattern} operator. It extends CEP to unstructured document streams by seamlessly fusing LLM-based event extraction with automaton-based temporal pattern matching.

\noindent \textbf{Model and Syntax.}
Unlike traditional CEP systems where events are emitted by instrumented infrastructure, the atomic unit of matching in VectraFlow is a \textit{semantic event}: a foundational fact extracted on demand from unstructured text by an LLM and represented as a typed, timestamped tuple with a concise semantic description usable in pattern guards. 
\noindent VectraFlow supports the following constructs: 
\begin{itemize}[leftmargin=*, itemsep=1pt, topsep=2pt]
    \item \textbf{Sequence} ($A \rightarrow B \rightarrow C$): events occur in order with relaxed contiguity.
    \item \textbf{Conjunction} ($A \wedge B$): sub-patterns match in either order within a window.
    \item \textbf{Disjunction} ($A \vee B$): either sub-pattern suffices.
    \item \textbf{Negation} ($\neg A$): bounded absence of an event type~\cite{agrawal2008efficient, flinkcep}.
    \item \textbf{Quantifiers} \texttt{times($n$)}: \texttt{one\_or\_more}, or \texttt{optional}.
    \item \textbf{Temporal Constraint} (\texttt{within} $\Delta t$): bounds elapsed time between the first and last match.
\end{itemize}
These constructs compose freely, enabling expressive rules such as a sequence with an embedded negation and a time bound, e.g., \(\texttt{within}(A \rightarrow \neg B,\ \Delta t)\).



\noindent \textbf{Execution Semantics.}
The \texttt{sem\_pattern} operator runs in two stages:
(1) an LLM extractor emits a typed, timestamped event stream keyed by entity; and
(2) an Nondeterministic Finite Automaton (NFA) rule detector evaluates CEP rules over per-entity event sequences.
The operator is \emph{stateful}, maintaining active partial matches across arrivals.
This design separates semantic interpretation from temporal validation.


Following the SASE execution model~\cite{wu2006sase, agrawal2008efficient} as adopted by FlinkCEP~\cite{flinkcep}, each rule is compiled into a shared NFA with three transition types: \textsc{Take} (consume), \textsc{Ignore} (bypass), and \textsc{Proceed} ($\varepsilon$-transition). Any incoming event satisfying an initial transition spawns a lightweight \emph{NFA instance} that shares the compiled automaton while maintaining isolated runtime state (current node, matched events, and elapsed window). This compilation directly captures two key semantics. Sequence patterns implement \emph{skip-till-any-match} contiguity via \textsc{Ignore} edges, allowing partial matches to bypass irrelevant interleaved events. Negation is compiled as a bounded stop-state: an instance survives non-forbidden events but is killed upon a forbidden one. To ensure well-defined evaluation, negated patterns are bounded by a \texttt{within} window (enforced at compile time), and an instance becomes a match only if it survives the full window without triggering the negative guard.


\noindent \textbf{Experimental Results.}
We evaluated \texttt{sem\_pattern} on 256 clinical notes from MIMIC-IV~\cite{johnson2020mimiciv} across five complex event patterns involving sequences, negations, and time-bounded conjunctions.
We compared four configurations. \emph{Baseline} incrementally aggregates incoming documents and prompts the LLM to determine, at each step, whether the temporal pattern is satisfied over the expanding narrative. \emph{Baseline (+~RAG)} augments each judgment with retrieved context to reduce the accumulated narrative length. \emph{\texttt{sem\_pattern}} instead extracts typed events from each document via LLM and delegates temporal reasoning to the NFA engine, decoupling semantic extraction from pattern matching. \emph{\texttt{sem\_pattern} (+~RAG)} further focuses each extraction call on relevant passages via retrieval before invoking the LLM.
Results are reported for GPT-4o-mini~\cite{achiam2023gpt4}, Qwen3-8B, and Qwen3-4B~\cite{yang2025qwen3}; GPT-4o-mini is deployed on a private Azure server for secure MIMIC-IV data processing.

\begin{table}[htbp]
\centering
\caption{Efficiency-accuracy tradeoffs for semantic CEP approaches.}
\label{tab:cross_model_performance}
\renewcommand{\arraystretch}{1.15}
\resizebox{\columnwidth}{!}{%
\begin{tabular}{@{} l r c c c @{}}
\toprule
\textbf{Method} & \textbf{Total Tokens} & \multicolumn{3}{c}{\textbf{Average F1 Score}} \\
\cmidrule(l){3-5}
& & \textbf{GPT-4o-mini} & \textbf{Qwen3-8B} & \textbf{Qwen3-4B} \\
\midrule
Full Context Baseline         & 14.6M$^*$     & 0.675          & OOM$^*$        & OOM$^*$ \\
Full Context Baseline (+ RAG) & 7.0M          & 0.787          & 0.721          & 0.812 \\
\texttt{sem\_pattern}         & 5.7M          & 0.844          & 0.814          & 0.791 \\
\texttt{sem\_pattern} (+ RAG) & \textbf{3.1M} & \textbf{0.848} & \textbf{0.862} & \textbf{0.822} \\
\bottomrule
\end{tabular}%
}
\\ 
\raggedright \scriptsize $^*$The full-context baseline exceeded the practical GPU VRAM limits of the local Qwen deployments.
\end{table}

Table~\ref{tab:cross_model_performance} shows that \texttt{sem\_pattern} consistently improves the token-accuracy tradeoff. Relying on a single stateless LLM call to parse accumulating text, the full-context Baseline suffers from severe context bloat (14.6M tokens) and triggers Out-Of-Memory (OOM) failures on local Qwen deployments. Even when GPU memory is not a bottleneck, direct LLM pattern judgment performs poorly (F1 = 0.675). RAG improves both experimental settings, yet \texttt{sem\_pattern} (+ RAG) consistently delivers the best efficiency and accuracy across all models. The results show that \texttt{sem\_pattern} attains a better balance between efficiency and accuracy than full-context prompting, allowing long-horizon pattern detection with fewer tokens and improved precision.

\section{Demonstration Setup}
VectraFlow exposes the full lifecycle of semantic stream processing, from natural language specification to operator execution and temporal pattern detection, through a set of tightly integrated interactive views. Together, these views allow users not only to execute queries but also to understand, debug, and refine long-horizon semantic pipelines in real time.

\begin{figure*}[t]
    \centering
    \includegraphics[width=0.32\linewidth]{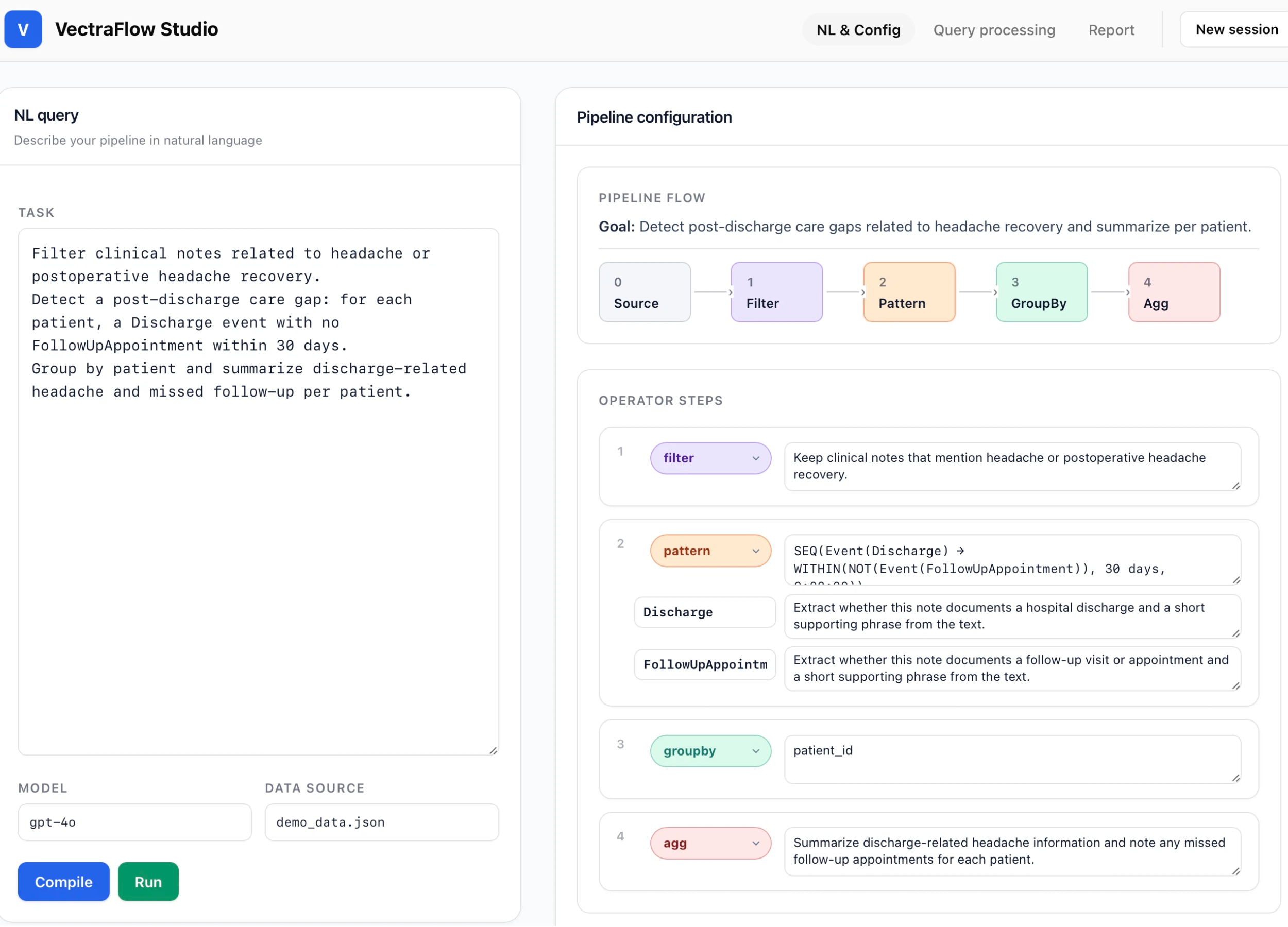}\hfill
    \includegraphics[width=0.32\linewidth]{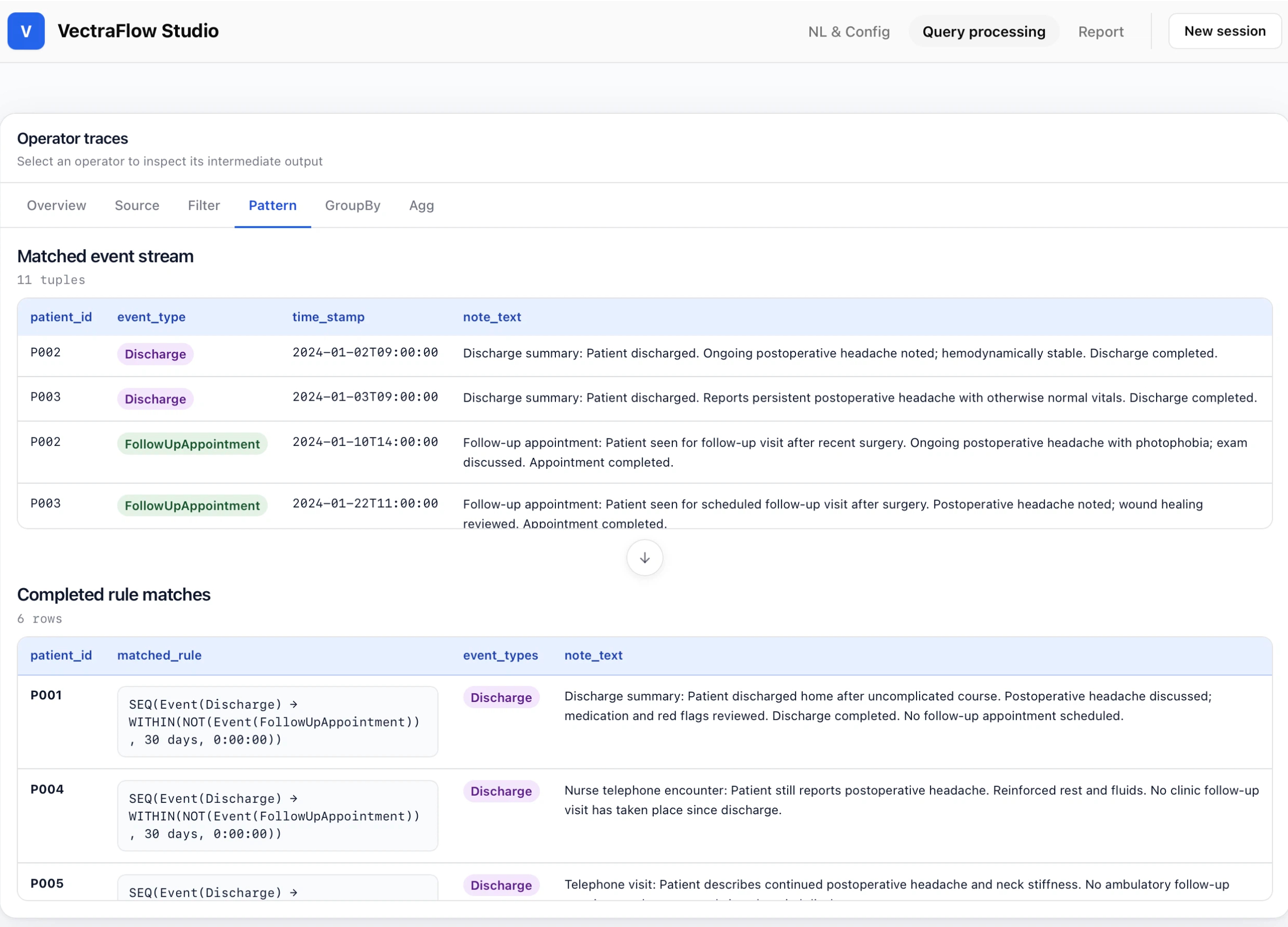}\hfill
    \includegraphics[width=0.32\linewidth]{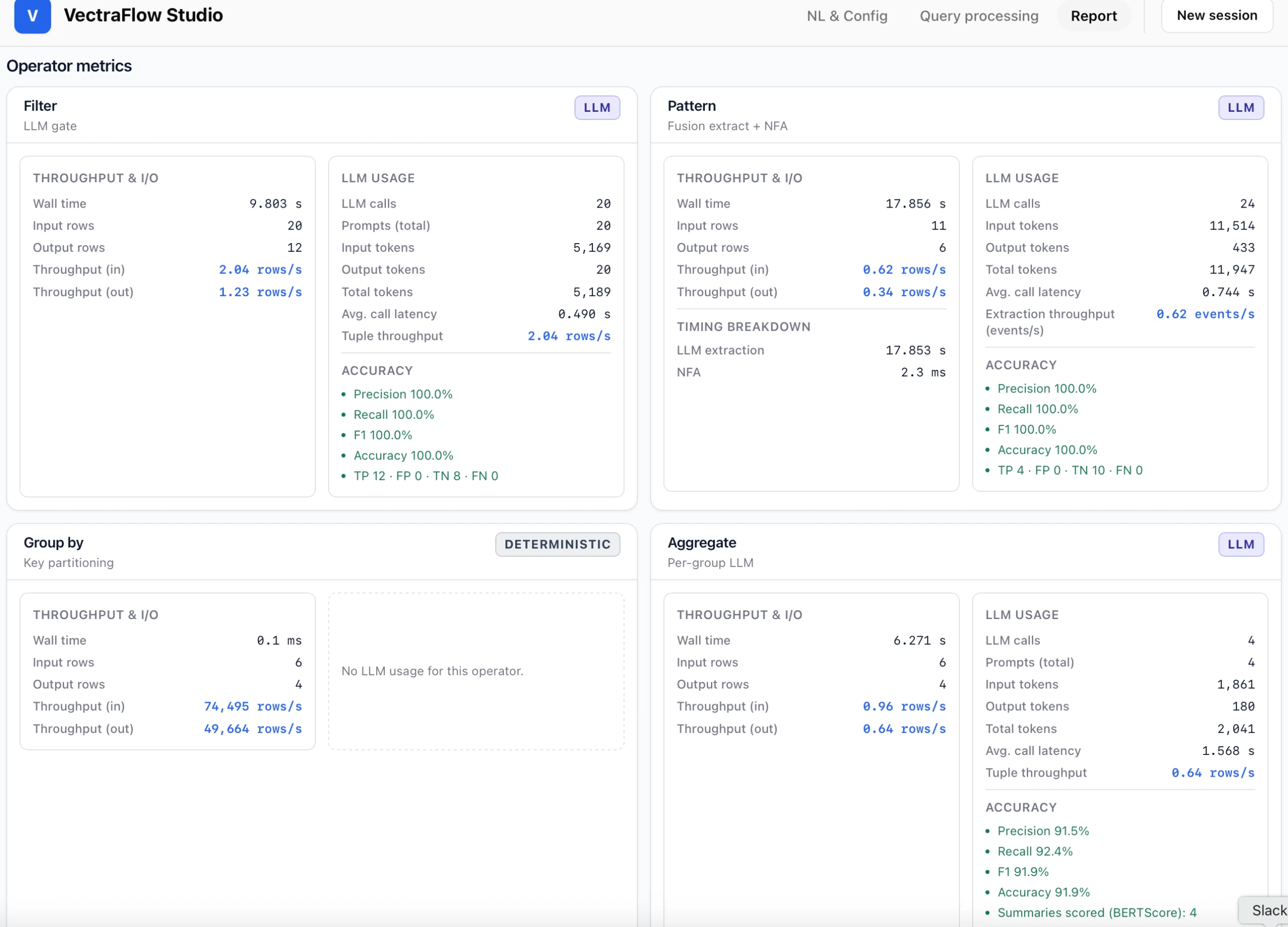}
    \caption{VectraFlow interactive interface. (a) \textbf{NL \& Config View}: Natural language to executable pipeline compilation. 
    \\(b) \textbf{Query Processing View}: Per-operator traces and intermediate outputs. 
    \\(c) \textbf{Report View}: End-to-end execution metrics and LLM profiling.}
    \label{fig:demo_views}
\end{figure*}

\noindent \textbf{Pipeline Authoring.}
Figure~\ref{fig:demo_views}(a) shows the NL \& Config view, where users describe their pipeline as a free-form natural language task over a synthetic clinical document stream derived from MIMIC-IV~\cite{johnson2020mimiciv}.
The pipeline synthesizer compiles this into an executable operator sequence in which each operator’s type and bound LLM instruction prompt are editable inline.
For \texttt{sem\_pattern}, compilation exposes two components: a formal pattern expression (e.g., \texttt{SEQ(Discharge, WITHIN(NOT(FollowUp), 30~days))}) and per-event LLM extraction prompts defining what the extractor looks for in each document.
Users can modify any step and recompile without restarting, enabling iterative refinement of both pipeline structure and event semantics. This tight edit-compile-execute loop enables rapid exploration of alternative semantic interpretations and pattern definitions, making the impact of prompt and operator changes immediately visible in downstream results.

\noindent \textbf{Pipeline Inspection.} Figure~\ref{fig:demo_views}(b) shows the Query Processing view, where intermediate results are inspectable at each pipeline stage via per-operator tabs.
The Pattern operator exposes two layers: the extracted event stream and the completed rule matches emitted by the NFA-based rule detector. For each emitted match, the system shows the matched pattern expression and supporting evidence span.
This makes the full transformation from unstructured documents to structured events to matched temporal patterns directly observable at every step. By exposing both semantic extraction and temporal reasoning stages, the system makes it possible to diagnose errors arising from either misinterpreted text or incorrect pattern logic, a distinction that is typically opaque in end-to-end LLM pipelines. This fine-grained visibility enables users to reason about the cost-accuracy tradeoffs of different operator configurations, and to identify bottlenecks arising from LLM invocation patterns or dataflow structure.

\noindent \textbf{Execution Profiling.}
Figure~\ref{fig:demo_views}(c) shows the Report view, presenting per-operator execution metrics across the full pipeline.
Every operator reports wall time, I/O row counts, and throughput; LLM-powered operators (e.g., Filter, Pattern) additionally surface token usage, call latency, extraction throughput, and accuracy metrics where ground truth is available.



\section{Conclusions}
VectraFlow introduces a unified framework for continuous semantic processing and event pattern detection over unstructured streams, combining LLM-based operators with stateful, long-horizon temporal reasoning. Its key innovations include a suite of continuous semantic operators and a semantic pattern operator that integrates event extraction with temporal rule matching within a single dataflow abstraction. This demo of VectraFlow enables users to compose, execute, and observe semantic pipelines over free‑form clinical document streams, exposing both intermediate operator behavior and end‑to‑end pattern detection.



\bibliographystyle{ACM-Reference-Format}
\bibliography{sample}

@misc{flinkcep,
  author       = {{Apache Flink}},
  title        = {{FlinkCEP} --- Complex Event Processing},
  howpublished = {\url{https://nightlies.apache.org/flink/flink-docs-release-1.20/docs/libs/cep/}},
  year         = {2024},
  note         = {Accessed: 2024}
}

@misc{esper,
  author       = {{EsperTech}},
  title        = {Esper Reference Documentation --- Event Pattern Operators},
  howpublished = {\url{http://esper.espertech.com/release-9.0.0/reference-esper/html/event_patterns.html}},
  year         = {2023},
  note         = {Accessed: 2024}
}

@misc{snowflake-match-recognize,
  author       = {{Snowflake Inc.}},
  title        = {{MATCH\_RECOGNIZE}: Snowflake Documentation},
  howpublished = {\url{https://docs.snowflake.com/en/sql-reference/constructs/match_recognize}},
  year         = {2024},
  note         = {Accessed: 2024}
}

@article{aurora,
  author  = {Daniel J. Abadi and Don Carney and U{\u{g}}ur {\c{C}}etintemel and
             Mitch Cherniack and Christian Convey and Sangdon Lee and
             Michael Stonebraker and Nesime Tatbul and Stan Zdonik},
  title   = {Aurora: A New Model and Architecture for Data Stream Management},
  journal = {The VLDB Journal},
  volume  = {12},
  number  = {2},
  pages   = {120--139},
  year    = {2003}
}

@inproceedings{wu2006sase,
  author    = {Eugene Wu and Yanlei Diao and Shariq Rizvi},
  title     = {High-Performance Complex Event Processing over Streams},
  booktitle = {Proceedings of the 2006 ACM SIGMOD International Conference on
               Management of Data},
  pages     = {407--418},
  year      = {2006}
}

@article{akdere2008plan,
  author  = {Mert Akdere and U{\u{g}}ur {\c{C}}etintemel and Nesime Tatbul},
  title   = {Plan-Based Complex Event Detection across Distributed Sources},
  journal = {PVLDB},
  volume  = {1},
  number  = {1},
  pages   = {66--77},
  year    = {2008}
}

@inproceedings{agrawal2008efficient,
  author    = {Jagrati Agrawal and Yanlei Diao and Daniel Gyllstrom 
               and Neil Immerman},
  title     = {Efficient Pattern Matching over Event Streams},
  booktitle = {Proceedings of the 2008 ACM SIGMOD International Conference
               on Management of Data},
  pages     = {147--160},
  year      = {2008}
}

@inproceedings{vectraflow-cidr,
  author    = {Duo Lu and Siming Feng and Jonathan Zhou and
               Franco Solleza and Malte Schwarzkopf and
               U{\u{g}}ur {\c{C}}etintemel},
  title     = {{VectraFlow}: Integrating Vectors into Stream Processing},
  booktitle = {Proceedings of the 15th Annual Conference on Innovative
               Data Systems Research},
  year      = {2025}
}

@article{vectraflow-cp,
  author  = {Shu Chen and Deepti Raghavan and U{\u{g}}ur {\c{C}}etintemel},
  title   = {Continuous Prompts: {LLM}-Augmented Pipeline Processing
             over Unstructured Streams},
  journal = {arXiv preprint arXiv:2512.03389},
  year    = {2025},
  url     = {https://arxiv.org/abs/2512.03389}
}

@inproceedings{toraman2024mide22,
  title     = {{M}i{D}e22: An Annotated Multi-Event Tweet Dataset for Misinformation Detection},
  author    = {Toraman, Cagri and Ozcelik, Oguzhan and Sahinu\c{c}, Furkan and Can, Fazli},
  booktitle = {Proceedings of the 2024 Joint International Conference on Computational Linguistics, Language Resources and Evaluation (LREC-COLING 2024)},
  year      = {2024},
  publisher = {ELRA and ICCL},
  pages     = {11283--11295},
  url       = {https://aclanthology.org/2024.lrec-main.986}
}

@article{patel2025semanticoperators,
  author    = {Patel, Liana and Jha, Siddharth and Pan, Melissa and Gupta, Harshit and Asawa, Parth and Guestrin, Carlos and Zaharia, Matei},
  title     = {Semantic operators and their optimization: Enabling {LLM}-powered analytics},
  journal   = {Proceedings of the VLDB Endowment (PVLDB)},
  volume    = {18},
  number    = {3},
  pages     = {4171--4184},
  year      = {2025},
  doi       = {10.14778/3749646.3749685}
}

@inproceedings{liu2025palimpzest,
  author       = {Liu, Chunwei and Russo, Matthew and Cafarella, Michael and Cao, Lei and Chen, Peter Baille and Chen, Zui and Franklin, Michael and Kraska, Tim and Madden, Samuel and Shahout, Rana and Vitagliano, Gerardo},
  title        = {Palimpzest: Optimizing AI-Powered Analytics with Declarative Query Processing},
  booktitle    = {Proceedings of the 15th Conference on Innovative Data Systems Research (CIDR)},
  year         = {2025},
}

@article{shankar2025docetl,
  author    = {Shankar, Shreya and Chambers, Tristan and Shah, Tarak and Parameswaran, Aditya G. and Wu, Eugene},
  title     = {DocETL: Agentic query rewriting and evaluation for complex document processing},
  journal   = {Proceedings of the VLDB Endowment},
  volume    = {18},
  number    = {9},
  year      = {2025},
  doi       = {10.14778/3746405.3746426}
}

@article{yang2025qwen3,
  author    = {Yang, An and Li, Anfeng and Yang, Baosong and Zhang, Beichen and Hui, Binyuan and Zheng, Bo and Yu, Bowen and Gao, Chao and Huang, Cheng and Lv, Chen and Zheng, Chen and Qiu, Zhenyu},
  title     = {Qwen3 Technical Report},
  journal = {arXiv preprint arXiv:2505.09388},
  year      = {2025}
}

@article{achiam2023gpt4,
  author  = {Achiam, Josh and Adler, Scott and Agarwal, Sandhini and Ahmad, Liane and Akkaya, Ilge and Aleman, Felipe L. and Almeida, Daniel and Altenschmidt, Johannes and Altman, Sam and Anadkat, Shantanu and Avila, Rafael},
  title   = {GPT-4 Technical Report},
  journal = {arXiv preprint arXiv:2303.08774},
  year    = {2023}
}

@misc{johnson2020mimiciv,
  author       = {Johnson, Alistair and Bulgarelli, Lucas and Pollard, Tom and Horng, Steven and Celi, Leo Anthony and Mark, Roger},
  title        = {MIMIC-IV},
  howpublished = {\url{https://physionet.org/content/mimiciv/1.0/}},
  note         = {Accessed: 2021-08-23},
  year         = {2020}
}

\end{document}